\documentclass[showpacs,amsmath,amssymb,floatfix,prb,twocolumn]{revtex4}
\usepackage{graphicx,amsmath}

\begin{document}

\title {Disorder induced transverse delocalisation in ropes of carbon nanotubes.}
\author{M.Ferrier}
\affiliation{Univ. Paris-Sud, CNRS, UMR 8502, F-91405 Orsay Cedex, France}
\author{A. Chepelianskii}
\affiliation{Univ. Paris-Sud, CNRS, UMR 8502, F-91405 Orsay Cedex, France}
\author{S.Gu\'eron}
\affiliation{Univ. Paris-Sud, CNRS, UMR 8502, F-91405 Orsay Cedex, France}
\author{H. Bouchiat}
\affiliation{Univ. Paris-Sud, CNRS, UMR 8502, F-91405 Orsay Cedex, France}

\begin{abstract}

A rope of carbon nanotubes is constituted of an  array of parallel  single wall  nanotubes with nearly identical diameters. In most cases the individual  nanotubes within a rope have different helicities and 1/3 of them are metallic. In the absence of  disorder within the tubes,  the intertube  electronic transfer is  negligeable because of the longitudinal wave vector mismatch between  neighboring tubes of different helicities. The rope can then be considered 
as a number of parallel independent ballistic nanotubes. On the other hand,  the presence of disorder  within the tubes favors  the intertube electronic transfer. This is first shown using a very simple model  where disorder is treated perturbatively inspired  by the work in reference \cite{maarouf00}.
 We then present numerical simulations on a tight binding model  of a rope.  Disorder induced  transverse delocalisation shows up as  a spectacular increase of  the sensitivity to the transverse boundary conditions  in the presence of small disorder. This is accompanied by an increase of the longitudinal localisation length. Implications on the nature of electronic transport within a rope of carbon nanotubes are discussed.
  
\end{abstract}
\maketitle

\section{Introduction}

 A rope  of  single wall carbon nanotubes (SWNT) is  generally made of ordered parallel  tubes with different helicities, but with a narrow distribution of diameters \cite{dresselhaus96,journet97}. The center of the tubes form a triangular lattice so that there is for each  metallic tube in a rope on average two neighboring tubes which are also metallic. In the absence of  disorder within the tubes,  the intertube 
 electronic transfer, defined as the matrix element of the transverse 
 coupling between two neighboring tubes, integrated over spatial coordinates, is 
 negligeable because of the longitudinal wave vector mismatch between 
tubes of different helicities \cite{maarouf00}. The rope can then be considered 
 as made of parallel independent  nanotubes. The transport is ballistic and the  dimension less conductance (in units of  the quantum conductance $G_Q =2 e^2/h$) is two times the number of metallic tubes within the rope. 
However, it has been shown \cite{maarouf00,tunney06} that  disorder 
 within the tubes favors intertube scattering by relaxing the  strict 
 orthogonality between the longitudinal components of the wave functions.  
 Using a very simple model  where disorder is treated perturbatively, 
  we  show in section II that   the intertube scattering 
 time is shorter than the elastic scattering time within a single tube.  In 
 tubes longer than the elastic mean free path,  this intertube scattering 
 can provide additional conducting paths to electrons which would otherwise 
 be localized in isolated tubes. In the  limit of localised transport along the tubes  we show  that the  longitudinal localisation length  is not a monotonous function of disorder and increases at moderate disorder. 
 In order to go beyond these  analytical results we have performed numerical simulations on a tight binding model  of coupled 1D chains with different longitudinal hopping energies.  This model described in section III mimics the physics of transport in a rope of carbon nanotubes in the sense that in the absence of disorder the electronic motion is localised within each chain. Transverse delocalisation as  a function of disorder is investigated through the sensitivity of eigen-energies to a change of transverse boundary conditions from periodic to antiperiodic.  
 
These results show that disordered ropes of carbon nanotubes can be 
 considered as anisotropic diffusive conductors, which in contrast to 
 individual tubes, exhibit a localization length that can  be much greater 
 than the elastic mean free path.

\section{Electronic structure of   ropes of carbon nanotubes}

\subsection{Band structure of a rope without disorder}
We consider a rope constituted from SWNT   with diameters ranging between 1.2 and 1.5 nm. It can be shown 
(see table \ref{diam}) that the tubes within such a rope can have  different kind of helicities.
Following the model developed by Maarouf and Kane \cite{maarouf00} 
one can characterize the electron wave functions at the Fermi energy $\epsilon_F$ with two  wave vectors $k_{\bot}$ and $k_{\parallel}$ respectively  perpendicular and parallel to the tube axis : $ |\Psi> = |k_{\bot}>|k_{\parallel}>$ such that  $\Psi (x,y) \sim e^{ik_{\bot}y}e^{ik_{\parallel}x}$. From this model, it is possible to compute  the matrix elements of the transverse coupling  hamiltonian $H_{\bot}$   between 2 tubes a and b:

\begin{equation}
\left\langle\Psi_a\left|H_{\bot} \right|\Psi_b\right\rangle = t_{\bot}(a,b) \delta_{k_{a\parallel},k_{b\parallel}}
\end{equation}
with $t_{\bot}(a,b)= t_T e^{-(1/4)Ra_0(k_{a\bot}-k_{b\bot})^2}$ and
 $t_T = t_G\sqrt{\frac{a_0}{4\pi R}} = 7.5\:$meV for an average tube radius $R = 0.7\:$nm, $t_G = 0.1\:$eV is the  inter-plane hopping energy  in graphite and $a_0=0.5\times10^{-10}m$ is the Bohr radius.
The term $\delta_{k_{a\parallel},k_{b\parallel}}$  is due to the othogonality between wave functions of different longitudinal  wave vectors and shows that tubes of different helicities are uncoupled to first order in $H_{\bot}$.  The exponential part takes into account  the mismatch between the $\pi$ orbitals belonging to neighboring  tubes of different helicities \cite{maarouf00}. 

\begin{figure}
\begin{center}
\includegraphics[clip=true,width=6cm]{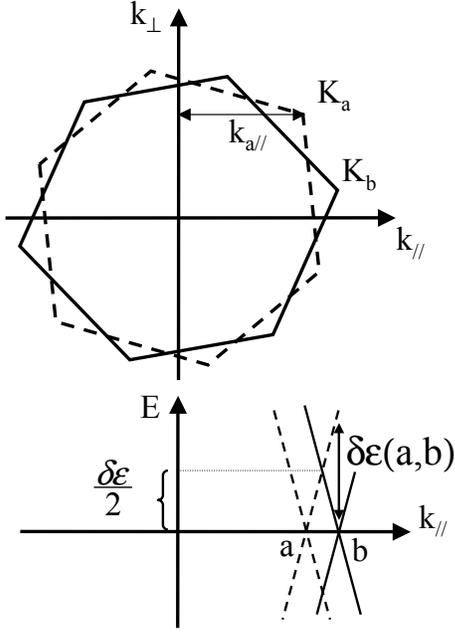}
%\leavevmode
%\epsfxsize=8cm

\caption{Top,  Brillouin zones of two  metallic tubes of different helicities.  Bottom:  Dispersion relation as  function of the longitudinal component of the wave vector  of the lowest energy bands for these 2 tubes for which $k_{a\parallel}^0  \neq k_{b\parallel}^0$ at the Fermi energy illustrating the second order coupling $\left\langle\Psi_a\left|H_{\bot} \right|\Psi_b\right\rangle = 2t_{\bot}(a,b)^2/\delta \epsilon(a,b)$ via the higher energy state  at energy $\epsilon_F+\delta \epsilon(a,b)/2$ for which  $k_{a\parallel}=k_{b\parallel}$.
\label{PZBchiral}}
\end{center}
\end{figure}
 Going beyond first order perturbation it is possible to show that neighbouring tubes of different helicities are  coupled  to second order in $H_{\bot}$ with a characteristic energy scale  $ e_{\bot 2}= 2t_{\bot}(a,b)^2/\delta \epsilon$ (Fig. \ref{PZBchiral}) involving  transitions to higher energy states for which $ \epsilon_b(k) = \epsilon_a(k) =\epsilon_F+\delta \epsilon_(a,b)/2 $. This  second order coupling gives rise to an intertube hopping  probability which is very small compared to the inverse ballistic time  $L/v_F$   of a  micron long  nanotube.
  It is thus reasonable to assume that, in a  rope constituted of tubes of different helicities,  electronic transport is confined within each ballistic tube and exhibits a strong 1D character.

\subsection{Disordered ropes in the perturbative regime} 
 In the presence of disorder,  plane waves  localised on  individual tubes  are  perturbed   into  wave packets and become much more sensitive to the transverse coupling leading to a  transport regime which is no longer
1D but can be delocalised on percolating  clusters  of metallic tubes within the rope.

 We consider  a very short range (on site) disorder potential, such that  its average $<V(x)>=0$ and  its variance $\sqrt{<V(x)^{2}>}=W$. 
\begin{equation}
V(x) = \sum_{a,x_a} W(x_a) \delta(x-x_a)
\end{equation}

where  the index  $a$  runs on the chains  constituting the rope, each of them  being characterized  by  its atomic sites $x_a$ and the  the  disorder potential $W(x_a)$ which  probability distribution  is given by:

\begin{equation}
\begin{array}{cccc}
P(W(x_a)) = &1/W&\rm{if}&w(x_a) \in [-W/2;W/2]\\
          &0  &\rm{otherwise}&  \\
\end{array}
\end{equation}
  
  The disorder perturbed  wave functions   can  be written to  first order in disorder: \par

\begin{equation}
|\Psi'> =|k_{\bot0}>\bigg(|k_{\parallel0}> + \sum_{i} \frac{\tilde{V}(k_{\parallel i}-k_{\parallel0})}{\epsilon(0)-\epsilon(i)}|k_{\parallel i}>\bigg) 
\end{equation}
where i runs over unoccupied states and $\tilde{V}(k)$ is the Fourier component of $V$ at wave vector k. We do not consider transitions involving different values of $k_{\bot}$ and neglect possible perturbation of $t_\bot(a,b)$ with disorder. 

This perturbed wave function contains  plane waves of all  values of parallel momentum.
As a result  electrons can hop from tube to tube and conserve their momentum. The  intertube coupling energy   between 2 tubes $a$ and $b$ is now $E_\bot = <\Psi'_a|H_\bot|\Psi'_b>$ which reads to first order in the disorder potential:
 
\begin{equation}
E_{\bot}(a,b) = <k_{a\bot}|H_{\bot}|k_{b\bot}> \frac2L\sum_{x_a}\frac{W(x_a)\cos((k_{\parallel a}-k_{\parallel b})x_a)}{\epsilon_b(k_{\parallel b}) -\epsilon_b(k_{\parallel a})}
\end{equation}  
The disorder average value of this coupling is zero but its typical value $
e_{\bot} = \sqrt{\overline{E_{\bot}^2}}$ is equal to:

\begin{equation}
e_{\bot}(a,b) = t_{\bot}(a,b)\frac{W}{3\delta\epsilon(a,b)}
\end{equation}  
with $\delta\epsilon(a,b)= \hbar v_F\left|{k_{\parallel a}-k_{\parallel b}}\right|$. 

\begin{figure}[hbt]
%\begin{center}
\includegraphics[clip=true,width=9cm]{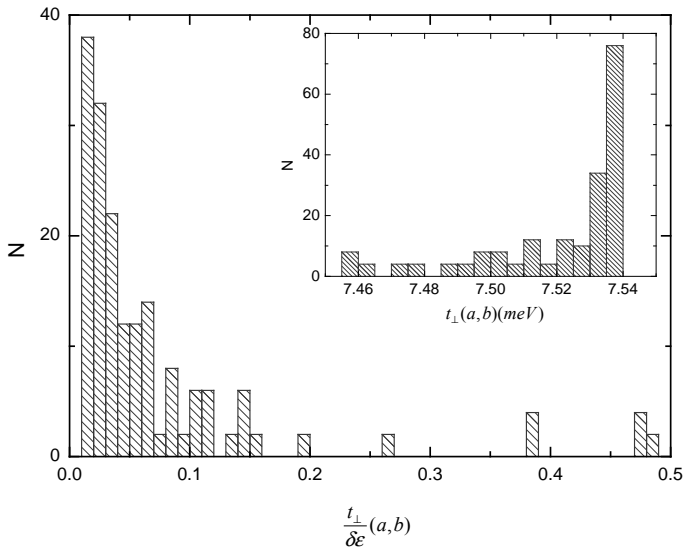}
%\leavevmode
%\epsfxsize=8cm

\caption{Histogram of possible values of the coefficients $t_{\bot}(a,b)$ and $\frac{t_{\bot}(a,b)}{\Delta\epsilon(a,b)}$ for all couple of helicities  $(a,b)$ of metallic tubes with a diameter between 1.2 and 1.5nm. The average yields $3 e_{bot} = 0.09$. 
\label{histo}}
%\end{center}
\end{figure}

This disorder induced  intertube coupling energy  is related to the second order coupling term calculated in the previous section through:  $e_{\bot}(a,b,W)/e_{\bot 2}(a,b)= W/t_{\bot}(a,b)$ which,  as will be shown below,  can be much larger than one  in a typical rope of SWNT.  The coefficients $t_{\bot}(a,b)$ and $\frac{t_{\bot}(a,b)}{\delta\epsilon(a,b)}$ were calculated for all the  helicities corresponding to metallic tubes given in table \ref{diam}  with diameter  between 1.2 and 1.5 nm, and are depicted in the histogram fig. (\ref{histo}). From these values we obtain the average value $ e_{\bot} = 0.03\:W$ .

\begin{table}[htb] 

\begin{tabular}{|c|c|c|c|c|c|c|c|}
\hline
helicity (n,m) & (9,9) & (10,10) & (11,8) & (12,9) & (12,6) & (13,7) & (15,6) \\ \hline
diameter (nm) & 1.24 & 1.38& 1.32 & 1.45 & 1.26 & 1.4 & 1.5  \\ \hline
helicity (n,m) & (14,5) & (13,4) & (16,4) & (15,3) & (17,2) & (16,1) & (18,0)\\ \hline
diameter (nm) & 1.36 & 1.23 & 1.46 & 1.33 & 1.44 & 1.32 & 1.43\\ \hline
\end{tabular}
\caption{Values of  possible helicities and diameters of  the 14 metallic tubes with a diameter  between 1.5 and 1.2 nm. Diameter and helicity  are related through: $D(n,m)=\frac{0.25}{\pi}\sqrt{m^2+mn+n^2}$.  There are also 24 insulating tubes in the same diameter range.   
\label{diam}}
\end{table}

%\begin{figure}
%\begin{center}
%\includegraphics[clip=true,width=7cm,angle=90]{edeW08.eps}

%\leavevmode
%\epsfxsize=8cm

%\caption{
%\label{DeltaE}}
%\end{center}
%\end{figure}

\begin{figure}
\begin{center}
\includegraphics[clip=true,width=9cm]{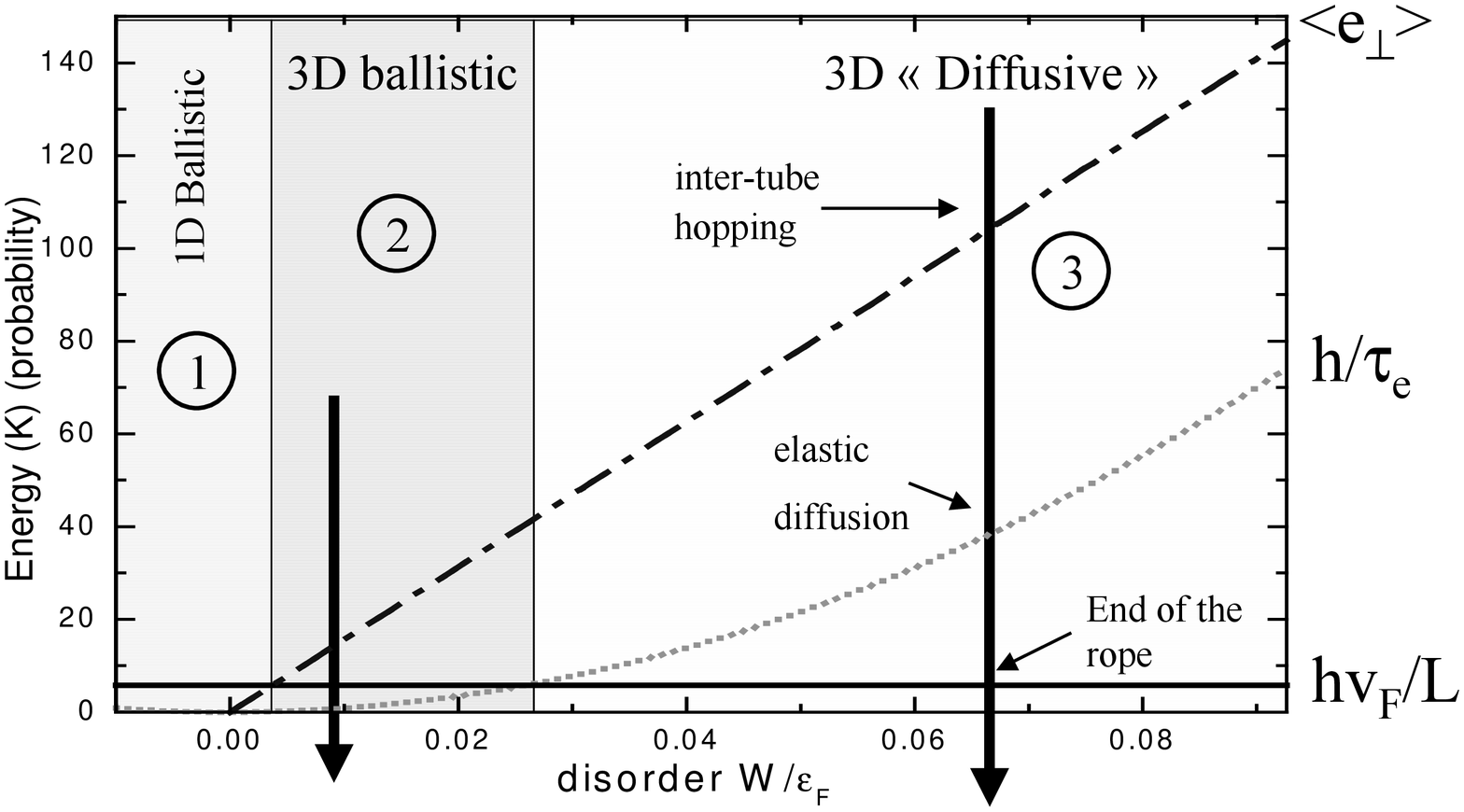}

%\leavevmode
%\epsfxsize=8cm

\caption{Different transport regimes depending on the amplitude of disorder compared in a rope of length $L =1\mu _m$ as discussed in the text. Region 1 $=$ 1D ballistic, region 2 $=$ 3D ballistic and region  3 $=$ 3D diffusive.  
\label{EdeW}}
\end{center}
\end{figure}
To investigate the nature of transport in ropes, it is interesting to compare the typical intertube hopping time $ \tau_h=\hbar/\left\langle e_{\bot} \right\rangle$   to the intra tube scattering time $\tau_e$ induced by the same disorder. The related elastic mean free path  $l_e =v_F \tau_e$ was calculated by  
White et Todorov \cite{white98} and found to be given by  $l_e = \frac{\epsilon^2_F}{W^2}n_C$ for a   tube  of $n_C$ carbon atoms along the circumference, where $\epsilon_F = 2.7\:$eV is the Fermi energy (measured from the bottom of the band)  and   $W^2$ is the variance of the  disorder (assumed to be short range). This value of  $l_e$ is  unusually large compared   to what is expected in an ordinary conductor, since it is proportional to the number of sites along the circumference of the tube. This is due to the existence of only 2 conduction modes at the Fermi energy regardless of the diameter of the tube.
As a result there exists a rather large range of disorder W for which $\tau_e$ is greater than $ \tau_h = \hbar/\left\langle e_{\bot} \right\rangle$,  which means that a charge carrier can visit several neighboring tubes between 2 elastic collisions. As shown schematically  on fig.\ref{EdeW} comparing the relevant time scales for  transport in a rope  of  micron length,  four different regimes  (1,2,3,4) can be reached as the amplitude of disorder is increased. 
\begin{enumerate}
	\item  At very low  amplitude of disorder, when both $\tau_e$ and $ \tau_h$ are  long compared to the  ballistic time $\tau_b= L/ v_f$,  transport is ballistic and one dimensional and electronic wave functions are localised within single tubes.
	\item When $\tau_h < \tau_b <\tau_e$  transport is still ballistic but wave functions are delocalised over several tubes.
	\item When  $\tau_b < \tau_h <\tau_e$   transport is diffusive along the percolating clusters of  metallic tubes along the rope as long  as the rope is shorter than the localisation length which typical  value is $\xi =2N_m l_e$; where $N_m$  is the number of metallic tubes in the largest  percolating cluster of metallic tubes within the rope. Typical values of $N_m$ for commonly investigated SWNT ropes  in experiments are given in the appendix.   
	\item At large disorder  $\tau_b < \tau_e <\tau_h$ electronic states  are localised within individual tubes  at the scale of $l_e$ (not shown in fig.\ref{EdeW}).
\end{enumerate}

From this qualitative model one expects that transverse transport but possibly also longitudinal transport is favored when increasing disorder in the rope. We present in the following section numerical simulations which confirm this statement.
     
%\begin{equation}

%\end{equation}
 \subsection{Analytical results in the localised regime}

In the  following we consider the  case where, in the absence of inter chain coupling,  electronic wave functions are localised within each tube  aligned along the x axis and  can be characterised  by the set of parameters $x_a$, $k_a$ and the localisation 
length $\xi_a$ such that:
\begin{equation}
\begin{array}{l}
	\Psi_a(x) =cos(k_a(x-x_a) \exp -( |x-x_a|/\xi_a)/\sqrt{ \xi_a} = \\  \Sigma_{k = 0}^{2n\pi/L} (f_a(k-k_a) + f_a(k+k_a) \exp\left[ i((x-x_a)\right]
\end{array}
\end{equation}
where $f_a$ is a Lorentzian function centered on $k_a$ of width $ \delta k_a = 2\pi/\xi_a$.  
In the presence of a small intertube coupling such as described by eq. \ref{Hperp}, one can easily compute the typical transverse coupling energy between two nearest neighbor tubes at lowest order in $t_ {\perp }$ :
\begin{equation}
\begin{array}{l}
<\Psi_a|H_{\perp}|\Psi_b> = \\\displaystyle  \frac {4t_ {\perp }(a,b)}{\lambda L}\Sigma_k  \frac{\exp ik(x_a-x_b)\lambda^4}{(\lambda^2+(k\pm k_a)^2)(\lambda^2+(k\pm k_b)^2)}
\end{array}
\end{equation}
 In the following we assume that  $ \delta k_a =\delta k_b = \lambda= 2\pi/\xi$ the summation $\Sigma_k$ runs on multiple values of $2\pi/L$  up to $n=L/a_0$ the number of sites on the tubes of length $L$.  Averaging over the random phase factors $\exp i(k-k')(x_a-x_b)$  for $k\neq k'$ leads to the average square of this coupling:
 \begin{equation}
 \begin{array}{l}
\Gamma_{ab}= |<\Psi_a|H_{\perp}|\Psi_b>|^2 = \\  4t_ {\perp }(a,b)^2/(\lambda L)^2  \displaystyle \Sigma_k  \frac{  \lambda^8}{\left[\lambda^2+((k\pm k_a)^2\right]^2\left[ \lambda^2+((k\pm k_b)^2 \right]^2}
\end{array}
\end{equation}
Keeping in the summation over $k$ only the $\lambda/(2\pi/L) $  terms centered around $k_a$ and $k_b$  within $\delta k = \lambda$,  finally yields:
\begin{equation}
\Gamma_{ab}= 4(t_ {\perp }(a,b)) ^2  \lambda L \displaystyle\frac{\lambda^2}{ \pi L^2(\lambda^2+(k_a-k_b)^2)^2}
\label{coupling}
\end{equation}

$\lambda$, the inverse typical localisation length for 1D disordered chains can be approximated by $\lambda = C a_0^{-1} W^2/E_F^2$ where the amplitude  W of the intra chain on site disorder is assumed to be small compared to the Fermi energy $E_F$. One can see from expression (\ref{coupling}) that the typical transverse coupling  between tubes of different helicities  $k_a \neq k_b$  obtained after averaging on the positions $x_a $ and $x_b$ increases with disorder like $W^6$ at low disorder and decreases  like $1/W^2$ at large disorder such that  $ \lambda  > |k_a-k_b|$.

\section{A simplified anisotropic tight binding model for a  rope of SWNT. }

\begin{figure}[htb]
\begin{center}
\includegraphics[clip=true,width=7cm]{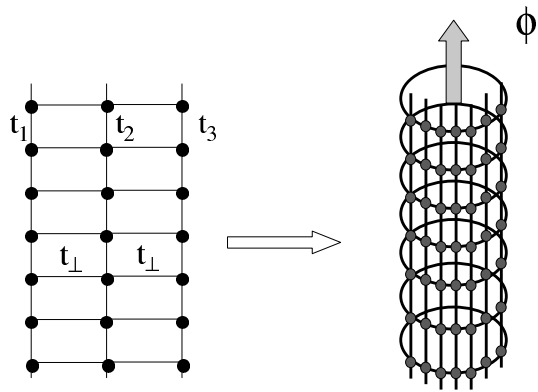}

%\leavevmode
%\epsfxsize=8cm

\caption{Left: Schematic modelisation of a rope of nanotubes by  $N_m$ slightly coupled tight binding chains  of different longitudinal energies $t_1 \neq t_2 \neq t_3$ etc....  Right: The chains  lye   on a cylinder threaded by a fictitious Aharonov Bohm flux which enables to investigate transverse transport.
\label{schema}}
\end{center}
\end{figure}

We investigate in the following a simplified tight binding model for the transport in a rope containing a  percolating cluster of $N_m$  metallic SWNT. As shown in the appendix, we expect that in commonly  investigated SWNT ropes $N_m$ is at most equal to 10. Each SWNT labeled $n$ is described by a simple 1D atomic chain  of $N_s$ sites with a nearest neighbor coupling energy  along the chain $t_n$ which can be different from one chain to the other.  The variables $t_n$ are randomly  distributed around their average value $t_{\parallel}$  with a square distribution of width $\delta t_ {\parallel }$. For convenience we take the chains  on the surface of a cylinder where only sites belonging to nearest neighbor chains are coupled (see Fig.\ref{schema}). This crudely reproduces the situation  of a hexagonally packed rope,  where each  metallic tube has on average 2 metallic  tubes as nearest neighbors.  The interchain transverse coupling is described by a transverse nearest neighbor coupling $t_ {\perp }\ll t_{\parallel}$.   The distribution among the $t_n$ plays the same role as the helicity distribution among the tubes in a rope i.e. the inter chain coupling becomes zero to first order in $t_ {\perp }$ since identical longitudinal Bloch wave vectors correspond to different energies $t_n \cos ka_0$,  except at half filling $ka_0 = \pi/2$. Note however that this disorder among the $t_n$ also implies a distribution of Fermi velocities which does not exist in CNT.  This leads to the following hamiltonian $H= H_{\parallel}+ H_{\perp }$.

\begin{equation}
H_{\parallel}= \sum _{n=1}^{ N_m}\sum_{n_s=1}^{N_s -1} 
t_n\left[|n, n_s><n,n_s +1|+ h.c. \right]
\end{equation}
 
\begin{equation}
\begin{array}{l}
H_{\perp}= t_{\perp }\sum_{n_s=1}^{N_s}{\large[}
  \sum _{n=1}^{N_m-1} |n, n_s><n+1,n_s | 
+ \\ 
\exp(i\phi)|N_m,n_s><1,n_s| 
 +h.c. {\large]}
\end{array}
\label{Hperp}
\end{equation}

The last term in $H_{\perp }$ corresponds to the periodic boundary conditions around the cylinder modeling the rope. This periodic boundary condition involves a phase factor $\exp(i\phi)$  equivalent
to a fictitious flux $\Phi$ through the cylinder  which modifies the phase $\phi=2\pi\Phi/\Phi_0$ of  transverse boundary conditions of the wave functions \cite{imry}.  Due  to the different values  of $t_n$  the wave functions are localised within each chain in the limit of very long chains $N_s\gg N_m$ and $t_ {\perp } \ll t_{\parallel}$. An extra disorder hamiltonian  $H_d$ is added either as a random distribution of on-site potentials $w _i$ of width $W$ or as an extra random  contribution      $\delta t_n(s, s+1)$  to the  nearest neighbor  coupling $t_n$  within the chain $n$ (bond disorder) characterized by a distribution of width $\delta t_w$  assumed to be independent of n.
\subsection{Numerical results.}

\begin{figure}[htb]
\includegraphics[clip=true,width=8cm]{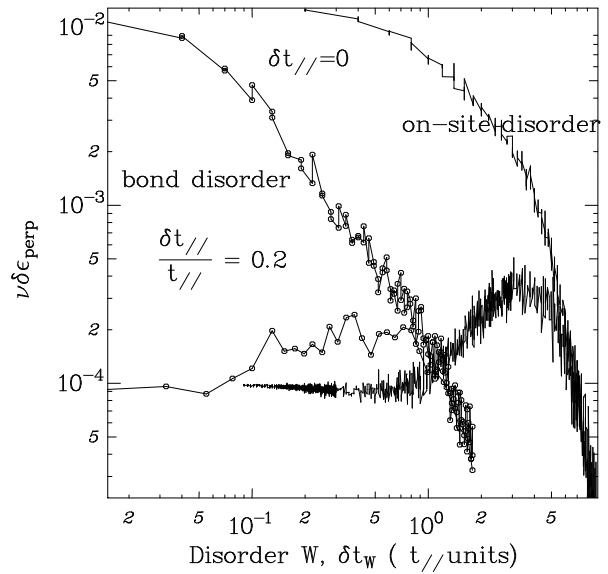}
\begin{center}		\caption{ Transverse delocalisation by disorder: Evolution of $\delta \epsilon_{perp} $ with the amplitude of disorder  for 10 coupled chains of 100 sites,   corresponding to  W for  on site disorder, and $\delta t_w$ for bond disorder. The situation  of identical values of $t_n$ corresponding to $\delta t_{\parallel} =0$  exhibits expected disorder induced localisation, whereas the situation with different values of $t_n$  where $\delta t _{\parallel} /t_{\parallel} =0.2$  shows a regime of disorder induced transverse delocalisation . The transverse hopping energy is chosen to be $ t_{\perp}= 0.06t_{\parallel} $. $\nu$ is the density of states (inverse nearest level spacing).
\label{comsitelien}}
\end{center}
\end{figure}

In order to investigate the interchain localisation and the effect of  disorder along the chains, we have calculated the eigenvalues of  $H(\phi) = H_0(\phi) +H_d$.  As shown in the seminal work of Thouless \cite{thouless72} the   sensitivity  of these eigenvalues to a change of the phase $\phi$ of the boundary conditions can be considered as a measure of the delocalisation of the wave functions on the various chains constituting the rope. More precisely we have computed the quantity $\delta \epsilon_{perp}=< |\epsilon _k(\phi=\pi) - \epsilon _k(0)|>_k$ where the average $<>_k$ is taken on  the $N_m N_s/4$  energy levels of the spectrum between 1/8 and 7/8 filling  excluding the region between 3/8 and 5/8 filling.   Around half filling the tight binding dispersion relations for all chains are  indeed crossing each other whatever the values of $t_n$ are and the model is inadequate.
For ropes of $N_m=10$ chains of $N_s=100$ sites the  phase dependent eigenvalues  were determined using a standard matlab  diagonalisation routine. On the other hand for longer ropes ($N_s=1000$) a Lancsos algorithm \cite{lancsos}  was used  to compute the eigenvalues spectrum.
The quantity $\delta \epsilon_{perp}$ calculated for ropes of ten tubes is shown on Fig.\ref{comsitelien} as a function of disorder strength  both for on site and bond disorder. 

\begin{figure}
\begin{center}
\includegraphics[clip=true,width=9cm]{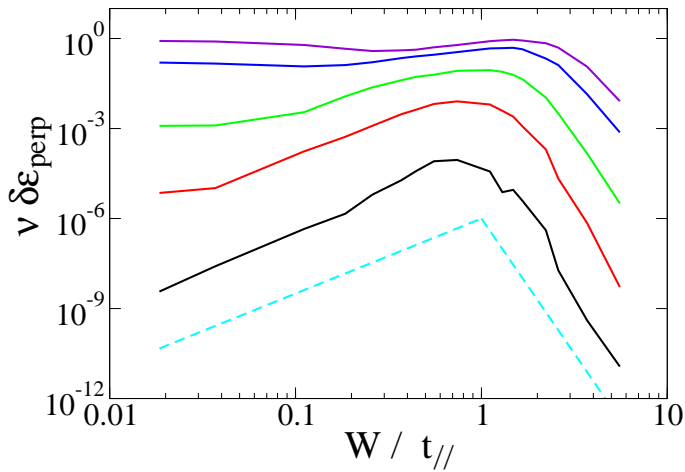} 
\end{center}
\caption{Results obtained  by diagonalisation of  a system 
 of $N_m = 10$ chains of $N_s = 1013$ sites.
Continuous curves, average energy difference $\delta \epsilon_{perp}$ as a function of on site disorder amplitude. 
From top to bottom: $t_{\perp} / t_{\parallel} = 0.1, 0.07, 0.03, 0.02, 0.01$. Note the  dependence in $t_{\perp}^10 $ at low disorder.
%(Data in /home/alik/Cprograms/almatr/swnt/Nx1013Ny10 ) 
Dashed curve power law fits approximation those curves with $2.5$ and $-9$ exponents (left and right part).
\label{figal1}}
\end{figure}

When the $t_n$ are all equal ($\delta t_{\parallel} =0$),  $\delta \epsilon_{perp }$ is a monotonously decreasing function of disorder as expected in the physics of standard localisation.
On the other hand when  the $t_n$ are different,   $\delta \epsilon_{perp}$ is very small at low values of disorder as expected, with  a power law dependence in $t_ {\perp }^{10}$ (this exponent corresponds to the hopping probability around a circumference with 10 sites), see Fig.\ref{figal1}. More important,  at fixed value of  $t_ {\perp }$,  $\delta \epsilon_{perp}$ increases with disorder amplitude,  goes through a maximum  and decreases  at large disorder.
As shown in  Fig.\ref{figal1} the conditions of observation of this non monotonous dependence of $\delta \epsilon_{perp }$ as a function of disorder amplitude depends drastically on the amplitude of the transverse hopping integral. It is clearly observed for very small values of $t_{\perp}$ with a low disorder increase in $W ^{2.5\pm0.5}$ followed by a decrease in $1/W^{9 \pm 1}$. This power law exponent  is consistent  with the analytical result derived in previous section for the quantity $\Gamma_{ab} ^{1/2} \propto 1/W$   describing the  large disorder hoping between two adjacent tubes, its extension   to hopping processes around a rope containing $N_m$ chains yielding a  decrease in $1/W ^(N_m)$.
 When increasing $t_{\perp}$ multiple order hopping  processes  in $t_{\perp}^{10}$  dominate the transverse transport and $\delta \epsilon_{perp}$ becomes independent of disorder  at low value of $W$ .

\begin{figure}
\begin{center}
\includegraphics[clip=true,width=9cm]{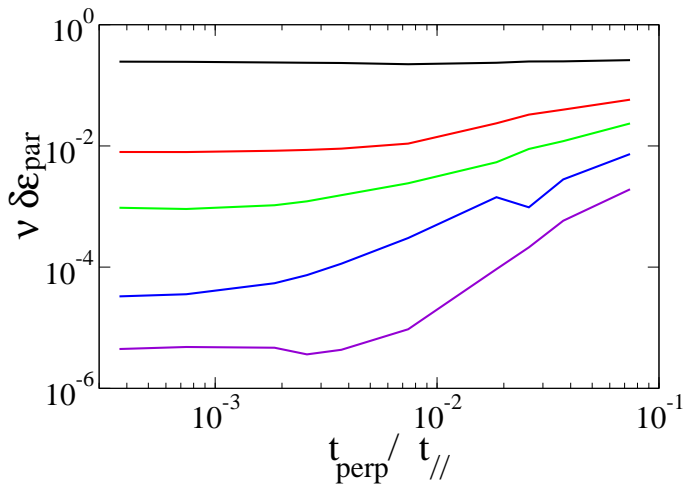} 
\end{center}
\caption{
Results obtained  by diagonalisation of  a system 
 of $N_m = 10$ chains of $N_s = 1013$ sites.
Average  typical energy difference $\delta \epsilon _{par}$  between the symmetric/antisymmetric boundary conditions in the longitudinal direction
as a function of $t_{\perp} /  t_{\parallel}$. 
From top to bottom $W /  t_{\parallel} = 0.55, 0.9, 1.1, 1.3, 1.5$
%(Data in /home/alik/Cprograms/almatr/swnt/Nx1013Ny10 )
\label{figal2} }
\end{figure}

The sensitivity to a phase shift along the chain direction was also investigated from the computation $\delta \epsilon _{par}=< |\epsilon _k(\phi_{\parallel}=\pi) - \epsilon _k(0)|>_k$  where $\phi_{\parallel}$ is the phase factor on periodic boundary conditions parallel to the tube axis. For long ropes (1000 sites along the longitudinal axis) it is possible to observe an increase of $\delta \epsilon _{par}$ with $t_{\perp}$, see 
 Fig. \ref{figal2}.   This behavior is associated with an increase of the longitudinal localisation length with $t_{\perp}$ in the range of disorder where the inter tube coupling increases with disorder. One can easily deduce from this figure that for the value of $W=1.3$
 the localisation length $\xi(t_{\bot})$ increases by  approximatively a factor 4 at $t_{\bot}/t_{\parallel} =0.06$. This is done assuming an $\exp(-L/\xi(t_{\bot}))$ behavior for $\delta \epsilon _{par}$.

%\begin{figure}
%\begin{center}
%\includegraphics[width= 0.8 \columnwidth,height=0.6\columnwidth,angle=0]{figalex1.eps} 
%\end{center}
%\label{wavefigal1}
%\caption{
%System size $N_x = 1013$, $N_y = 10$.
%Average energy difference $\delta \epsilon$  between the %symetric/antisymetric boundary conditions in the $x$ direction
%as a function of disorder amplitude. 
%From top to down  at $W / < t_x > = 1$ : $t_y / < t_x > = 0.1, 0.07, 0.03, 0.02, 0.01$. 
%(Data in %/home/alik/Cprograms/almatr/swnt/Nx1%013Ny10 ) 
%}
%\end{figure}

\section{Conclusion: implication for the transport in ropes on carbon nanotubes.}

We have shown that a rope of single wall carbon nanotubes of different helicities is expected to exhibit very different regimes of electronic transport depending on the amount of disorder. When disorder is very small electronic transport takes place in independent 1D ballistic tubes. The conductance is then expected to be $G=2NG_Q$ where N is the number of connected metallic tubes. The experimental observation of a strong shot noise reduction in short  low resistive ropes  \cite{roche02} confirms the ballistic nature of transport in these ropes. 
On the other hand disordered tubes are expected to behave as  3D diffusive multi channel conductors whose  maximum value of conductance is $G=2N_m(l_e/L)G_Q$ when L is smaller than the localization length $\xi = 2N_ml_e$, where $N_m$ is the number of metallic tubes in the largest percolating cluster of metallic tubes in the rope. Four probe transport measurements  on  NT ropes after ion irradiation damage  \cite{avouris00}  have been shown to give information on the intertube hopping processes within the rope which  was found to increase with disorder. These findings  can   also  be  interpreted  taking into account the increase of the intertube scattering rate with disorder. 
The physics of intertube transfer has also been shown to play an important role in the transport in multiwall nanotubes \cite{bourlon04,revueNTcharlier07}. The situation is however different than in  ropes since each tube is only coupled within first order to two other tubes (nearest inner and outer shells)  and  moreover only the most external tube is connected to electrodes.

Let us also mention that a  similar scenario of disorder induced delocalisation has been predicted to take place in networks of disordered polymers as discussed in \cite{prigodin93,zambetaki96,dupuis97}.

 We finally note that these different types of transport are also expected to influence the  superconductivity   observed at very low temperature on ropes of carbon nanotubes.
The superconductivity  in weakly disordered ropes has been observed to exhibit  a strongly 1D character with a T=0, H=0 transition. On the other hand more resistive ropes
exhibit a   broad transition  at finite temperature characteristic of a  multi channel quasi 3D system \cite{kociak01,kasumov03}. 

Acknowledgments: We acknowledge very fruitful discussions with Christophe Texier, Nicolas Dupuis, Michael Feigelmann,  Piotr Chudzinski  and Alexei Ioselevitch on this problem.

\section{Appendix: Determination of the typical size of percolating cluster of metallic tubes within a rope.}
\begin{figure}
\begin{center}
\includegraphics[clip=true,width=9cm]{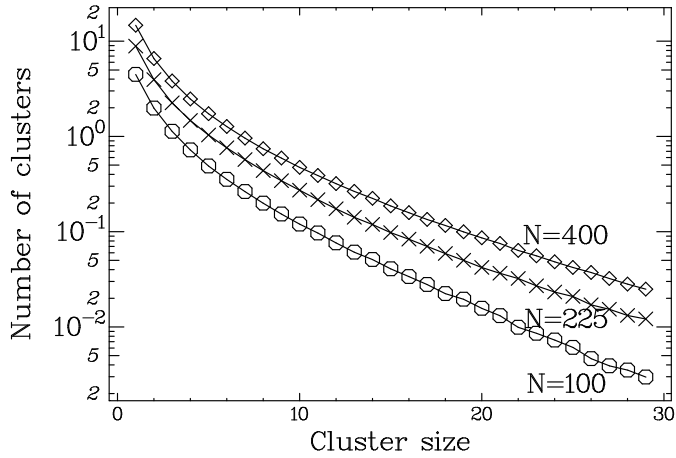} 
\end{center}
\caption{Size distribution of the number of tubes n in a percolating  cluster of metallic tubes for ropes containing N=100 and N=400 tubes among which 1/3  on average are  metallic. 
\label{clustersize} }
\end{figure}

  Carbon nanotubes in a rope are arranged according to a triangular network, with on average 1/3 of  metallic tubes  and 2/3 of semiconducting ones. Since 1/3 is below the percolating  threshold, 1/2, for nearest neighbor  couplings  in the 2D triangular lattice, the metallic tubes do not percolate over the full rope and constitute disconnected clusters which size distribution depends on the number of tubes  within the rope. We have calculated  numerically this size distribution for ropes containing a few hundred tubes. This result is shown on Fig.\ref{clustersize} for ropes containing 100 and 400 tubes.  The size distribution decays exponentially with the number of tubes in a cluster.   We find that for a rope containing 400  tubes the average cluster size is 3 but by integrating the number of clusters of size above a given value we find that  there is at least  one  cluster of tubes of size above 15. 
  
\end{document}